# Origin and Control of Polyacrylonitrile Alignments on Carbon Nanotube and Graphene Nanoribbon

Juho Lee,[1] Ji Il Choi,[1] Art E. Cho,[2] Satish Kumar,[3] Seung Soon Jang,[3] and Yong-Hoon Kim[1,*]

[1]Graduate School of Energy, Environment, Water, and Sustainability, Korea Advanced Institute of Science and Technology, 291 Daehak-ro, Yuseong-gu, Daejeon 305-701, Korea.

[2]Department of Bioinformatics, Korea University, Sejong, 339-700, Korea.

[3]School of Materials Science and Engineering, Georgia institute of Technology, 771 Ferst Drive, N.W., Atlanta, GA, USA

While one of the most promising applications of carbon nanotubes (CNTs) is to enhance polymer orientation and crystallization to achieve advanced carbon fibers, the successful realization of this goal has been hindered by the insufficient atomistic understanding of polymer-CNT interfaces. We herein theoretically study polyacrylonitrile (PAN)-CNT hybrid structures as a representative example of polymer-CNT composites. Based on density-functional theory calculations, we first find that the relative orientation of polar PAN nitrile groups with respect to the CNT surface is the key factor that determines the PAN-CNT interface energetics and the lying-down PAN configurations are much more preferable than their standing-up counterparts. The CNT curvature is identified as another important factor, giving the largest binding energy in the zero-curvature graphene limit. Charge transfer analysis explains the unique tendency of linear PAN alignments on the CNT surface and the possibility of ordered PAN-PAN assembly. Next, performing large-scale molecular dynamics simulations, we show that the desirable linear PAN-CNT alignment can be achieved even for relatively large initial misorientations and further demonstrate that graphene nanoribbons are a promising carbon nano-reinforcement candidate. The microscopic understanding accumulated in this study will provide design guidelines for the development of next-generation carbon nanofibers.

# 1. Introduction

Thanks to the outstanding mechanical and thermal properties with the high specific strength, carbon fibers are ideally suited for applications in diverse high-technology sectors such as automotive, aerospace, and nuclear engineering.[1-3] Carbon fibers are produced from polymer precursors through a series of spinning, stabilization, and carbonization processes,[2, 4] and a key obstacle in achieving high-quality carbon fibers is the presence of amorphous regions within the polymer matrixes that result in defects within the synthesized carbon fibers.[1] A promising strategy to improve order in polymers is to introduce carbon nanotubes (CNTs) as a template for polymer alignment and orientation and as a nucleating agent for polymer crystallization.[1, 3, 5, 6] Previous studies indeed reported that the use of small amount of CNTs notably improves the crystallization of polymers as well as the mechanical properties and chemical inertness of carbon fiber products.[7, 8] Importantly, the electrical and thermal conductivities of carbon fibers can be increased with the incorporation of CNTs into polymers, making it also very relevant for energy and flexible electronics applications.[5, 9] While the alignment geometry of CNT-polymer interface is apparently a key factor that affects the characteristics of produced fibers,[1, 3, 6] the molecular-level understanding of the polymer-CNT interface is still lacking, making further systematic improvements difficult.

Herein, applying a multi-scale approach that combines first-principles density functional theory (DFT) calculations and force-field (FF) molecular dynamics (MD) simulations,[10] we unravel the unique electronic and atomistic characteristics of polyacrylonitrile-CNT interfaces. Polyacrylonitrile (PAN) is currently the most common carbon fiber precursor and accounts for more than 90 percent of all carbon fiber production.[11] It is also one of the most promising carbon fiber precursors that can be utilized with CNTs.[7] Regarding the critical interface geometry in high-performance PAN-CNT composite fibers, the stacking of linearly aligned PANs along the CNT axis or the development of transcrystallinity was experimentally observed.[8, 12] Although this linear alignment tendency of PANs on the CNT surface was reproduced in an MD study,[13] the mechanisms behind this trend are not yet clearly understood. In general, while modeling and simulation can play an important role in the development of next-generation carbon fibers by providing theoretical understanding at the atomic level,[13, 14] they still remain in an early stage and particularly most previous works were limited to the classical molecular mechanics and dynamics level.

Our DFT calculations identify the preferable atomic configurations for PAN-CNT binding, and further show that the binding strength can be increased with the decrease in CNT curvature. The quantum mechanical nature of these simulations also establishes the connection between structural properties of PAN-CNT interfaces with their electronic structure: By analyzing the charge transfer characteristics for the representative PAN-CNT configurations, we observe the appearance of linearly extended side-by-side electron accumulation and depletion regions on the CNT surface. This explains not only the unique tendency of linear alignment of PANs on CNT with strong adhesion strength, but also via the compensation between positive and negative charge transfers at the PAN-CNT interface region the possibility of uniform dispersion of CNTs within the PAN matrix and improved large-scale PAN-PAN assembly. Based on the DFT-derived information on the energetic and electronic properties of PAN-CNT interfaces, we next carry out FF MD simulations to study the large-scale PAN-CNT conformations in the presence of multiple PAN chains. We especially demonstrate that graphene nanoribbions (GNRs) exhibit a very strong propensity to induce linear alignments of PANs adsorbed on them, and suggest that GNRs as a promising carbon nano-reinforcement candidate.

# 2. Results

In **Figure 1**a, we first show the three representative configurations of PAN adsorbed on (5,5) single-walled CNT, which in terms of the radial orientation of the polar PAN nitrile group (−C≡N:) can be categorized as one lying-down **L1** and two standing-up **S1** and **S2** configurations. Note that in the latter set, the PAN nitrile groups in **S1** and **S2** are pointing outward from and inward toward the CNT axis, respectively (For more detailed categorization and discussion of PAN-CNT conformations, see Supporting Information **Figure S1** and **Table S1**). Taking three PAN repeat units (RUs) for each case, we additionally scanned the energetic variations with respect to the relative position of PAN on the CNT surface (see Methods and Supplementary Information **Figure S2**): In terms of the location of the nitrogen atom in the head of the middle PAN RU in the configurations **L1** and **S2** or the corresponding hydrogen atom in the configuration **S1**, the on-top, hollow, and bridge sites of the (5,5) CNT surface were compared (**Figure 1b**). For all the angle-site combinations, we calculated the PAN-CNT binding energies according to

$$E_b = E_{\text{PAN-CNT}} - (E_{\text{CNT}} + E_{\text{PAN}})/NRU, \quad (1)$$

where $E_{\text{PAN-CNT}}$, $E_{\text{CNT}}$, and $E_{\text{PAN}}$ are the energies of the PAN-CNT complex, CNT, and PAN, respectively, and $NRU$ is the number of repeat units (three). The reliability of these computational results was first confirmed by employing several DFT exchange-correlation functionals (for the details, see Methods and Supplementary Information Table S2), and all calculations were finally performed within DFT-D3.[15] Overall, we found that the PAN-CNT binding strength is mainly determined by the adsorption angle of PAN on the CNT surface, while it is minimally affected by the binding site (Supplementary Information Table S3). Defining the radial orientation angles $\theta$ of the PAN nitrile group ($\overrightarrow{CN}$) with respect to the CNT radial direction ($\overrightarrow{OV}$), the optimized **L1**, **S1**, and **S2** configurations gave 100.7°, 31.9°, and 156.5°, respectively. Then, the binding energy of the lying-down PAN configuration **L1** was −0.197 eV·RU$^{-1}$. On the other hand, that of the standing-up configuration **S1** was slightly reduced to −0.169 eV·RU$^{-1}$, and the standing-up configuration **S2** (Figure 1a) turned out to be energetically least favorable with the binding energy of −0.103 eV·RU$^{-1}$.

Next, we discuss the curvature of CNT surface as another important factor that affects the PAN-CNT binding trend. The influence of nanoscale curvature of CNTs on molecular adsorption has been extensively studied in the context of various applications such as coating,[16] bio/chemical sensor,[17] hydrogen storage,[18] DNA sequencing,[19] etc. To analyze the CNT curvature effect, in addition to the (5,5) CNT (diameter $D$ = 6.78 Å) employed above, we prepared (10,0), (9,0), (10,10), and (19,0) CNTs with $D$ = 7.83 Å, 7.05 Å, 13.56 Å, and 14.87 Å, respectively, as well as graphene as the ideal infinite-diameter-limit system. Note that, by comparing the semiconducting zigzag CNT cases with the similar-diameter metallic armchair CNT ones, i.e. (10,0) vs. (5,5) CNTs and (19,0) vs. (10,10) CNTs, we can also check the influence of the CNT metallicity on the PAN-CNT binding. As summarized in **Figure 2**a (see also Supporting Information **Tables S4**), considering the PAN adsorbed at the most-preferable hollow binding sites, we observed that binding strength increases with the decrease in CNT curvature. Thus, the zero-curvature graphene case within the configuration **L1** exhibits the strongest PAN-graphene binding energy of $E_b$ = −0.263 eV·RU$^{-1}$. Although the PAN-CNT binding strength for the metallic armchair CNTs is slightly higher than that for semiconducting zigzag CNTs, we conclude that the effects of CNT chirality or metallicity are overall weaker than that of CNT curvature. Additionally, note that the binding-energy ordering of **L1** > **S1** > **S2** is not altered with the variation in CNT diameter and the ordering tendency is further strengthened with the decrease in CNT curvature (Figure 2a). Note that the strong PAN-CNT binding dominated by the **L1** configuration fulfills a precondition to achieve a high degree of CNT dispersion. Based on FF MD simulations, we will later show that these atomistic details are reflected on the large-scale PAN alignments and further provide the design rules for the optimization of PAN-CNT interfaces.

As discussed earlier, one of the most intriguing aspects of the PAN-CNT interface that will be the key to understand its excellent mechanical properties is the unique tendency of linear alignment of PAN on the CNT surface observed in experiments.[8, 12] While this behavior was reproduced in a previous FF MD simulation study,[13] its atomistic origins are not yet understood. To explain the mechanisms of such PAN-CNT binding behavior in terms of the PAN-CNT interfacial electronic structure, using the energetically most favorable configuration **L1**, we calculated the charge density differences,

$$\Delta\rho = \rho_{\text{CNT-P}} - (\rho_{\text{CNT}} + \rho_{\text{PAN}}), \qquad (2)$$

where $\rho_{CNT-PAN}$, $\rho_{PAN}$, and $\rho_{CNT}$ are the charge densities of the PAN-CNT composite, PAN, and CNT, respectively. As shown in **Figure 2c**, we observed that significant electron depletion with the maximum value ranging between −0.054 $|e|$·atom$^{-1}$ for the (9, 0) CNT and −0.101 $|e|$·atom$^{-1}$ for graphene is induced on the CNT/graphene region in contact with the highly polar PAN nitrile groups (more specifically, N lone-pair orbitals). At the same time, large electron accumulation with the maximum value ranging between +0.044 $|e|$·atom$^{-1}$ for the (9, 0) CNT and +0.099 $|e|$·atom$^{-1}$ for graphene was induced along the CNT/graphene region where the PAN backbone was placed. The locally large amount of charge transfer clarifies the electronic origin of strong PAN-CNT biding. Moreover, the side-by-side linearly extended configuration of the positive and negative charge transfers provides the atomistic explanation of the (Coulomb repulsion-driven) propensity of linear PAN alignment on CNT. Also of note is that the amounts of accumulating and depleting charges are almost equivalent, making the net charge transfer between CNT and PAN very small (about −0.04 $|e|$·RU$^{-1}$ Bader charge; Figure 2b). We point out that these unique charge transfer features have significant implications for the large-scale PAN assembly and ordering: As shown in **Figure 2d (left panel)**, due to the negligible net PAN-CNT charge transfer, the long-range (near the top and bottom box boundary regions) electrostatic potential induced by Δρ becomes negligible in the **L1** configuration. This is in clear contrast with, e.g., the **S1** interface configuration that develops significant long-range electrostatic potential profile (**Figure 2d right panel**). Here, we emphasize that the latter (non-counterbalancing interface charge transfer and resulting large long-range electrostatic potential fluctuations) corresponds to the situation of typical polymer-CNT interfaces. Namely, the unique PAN-CNT interfaces dominated by the **L1** configuration will not act as charged defect centers and be able to promote the long-range ordered PAN-PAN stacking. The enhanced transcrystallinity will then eventually achieve the overall good load transfer within the composites.

To additionally examine how the atomistic details of PAN-CNT interfaces characterized in DFT studies are reflected on the macroscopic conformations of PAN-CNT composites, we next carried out large-scale FF MD simulations with a Dreiding FF optimized for graphite [20] and (reflecting the negligible net PAN-CNT charge transfer seen in DFT calculations) zero atomic charges on CNT. We carefully tested the FF parameters against the DFT results discussed above and particularly confirmed that the binding energy trend of **L1** > **S1** > **S2** is well reproduced (For details, see Supporting Information **Table S5**).

We first discuss the single PAN chain MD simulation results. For the (5,5) CNT and 100 RU PAN models, we tested several different initial PAN orientations with respect to the CNT axis and compared eventual binding conformations. In **Figure 3**a, we show the final MD geometries obtained for the initial axial tilt angles $\phi$ (defined as the angle between the end-to-end vector of stretched polymer and the CNT axis) of 0°, 30°, and 45° (see also Supporting Information **Figure S3a**). The degree of linearity of a PAN adsorbed on CNT in our simulations can be quantified by measuring the PAN adsorption length, which is defined as the distance between the uppermost and lowermost constituent atoms in the PAN chain adsorbed along the CNT axis. We observe that, while rather long adsorption lengths of 12.97 and 11.50 nm were induced for the case of $\phi = 0°$ and 30° cases, respectively, an additional increase of PAN tilt angle to 45° resulted in the agglomeration of PAN and accordingly a drastically reduced adsorption length of 5.09 nm. We found that the count of PAN nitrile groups in contact with CNT or the quality of the PAN-CNT interface increases with the PAN adsorption length: The number of nitrile group at the PAN-CNT interface were 91 and 85 RUs for the initial axial tilt angle $\phi = 0°$ and 30° cases (**Figure 3b**; see also Supporting Information **Figure S4**), respectively. While almost all RUs are adsorbed on CNT in these low $\phi$ cases, the number was much reduced to 59 RUs for the $\phi = 45°$ case. Based on the nitrile groups counted at the interface, we further analyzed the distribution of the radial orientation angle $\theta$ of the nitrile group. The results show that the majority of RUs adopt $\theta = 100°\sim110°$, indicating that the configuration **L1** is dominant at the PAN-CNT interface (**Figure 3c**). The **S1** and **S2** configurations followed in terms of the frequency, which confirms that the large-scale PAN-CNT binding trend faithfully reflects the PAN-CNT binding energy ordering of **L1** > **S1** > **S2** identified in DFT calculations (Figure 2a).

We finally consider the more realistic situation where multiple PAN chains are present around CNTs, or include steric hindrance as well as PAN-PAN inter-chain interaction effects. Especially, with the purpose of discovering design rules to improve the transcrystallinity and in view of the CNT curvature-dependent PAN binding characteristics identified in DFT calculations, we considered (5,5) and (10,10) CNTs as well as zero-curvature graphene nanoribbon (GNR) as the nucleation sites for PANs. As the GNR model, we employed a hydrogen edge-terminated zigzag GNR composed of six zigzag chains (width ~ 13.6 Å). In addition, to fix the density of PANs at the CNT or GNR interfaces approximately identical in the three cases, we included four PANs for the (5,5) CNT and GNR cases and eight PANs for the (10,10) CNT counterpart (see Supporting Information **Figure S3b** and **Table S6**). Furthermore, we chose a rather large initial axial tilt $\phi = 45°$ to accommodate the possibility of generating amorphous conformations of adsorbed PANs. The simulation results are summarized in **Figure 4**a. First of all, we can immediately notice that the degradation of linear alignment tendency at $\phi = 45°$ in the single PAN case (Figure 3a) disappeared. For example, the adsorption length of multiple PAN chains in the (5, 5) CNT case reached 10.44 nm, much longer than that of the single chain case (5.09 nm). We thus conclude that the crowding of multiple PANs or steric hindrance around CNTs or GNRs could be beneficial in enhancing the transcrystallinity.

We now compare the (5,5) CNT, (10,10) CNT, and GNR cases. To properly quantify their transcrystallinity, we measured in addition to the PAN adsorption lengths the cylindrical distributions of PAN nitrile groups according to

$$\rho_{C\equiv N}(r) = 2\pi r \rho \delta r / A, \qquad (3)$$

where $\rho$, $r$, and $A$ are the number density of nitrile groups, distances of nitrile groups from the CNT axis, and circumferential area of CNT, respectively. For the GNR case, we measured the rectangular distribution of PAN nitrile groups around GNR (For details, see Supporting Information **Figure S5**) The comparison clearly shows that the PAN transcrystallinity increases with the decrease in CNT curvature: The adsorption length increases from 10.44 nm for (5,5) CNT to 11.82 nm for (10,10) CNT and to 15.18 nm for GNR. Correspondingly, the number density of nitrile groups in the interface region increased from 8.83 counts·Å$^{-1}$ for (5,5) CNT to 9.41 counts·Å$^{-1}$ for (10,10) CNT and to 9.74 counts·Å$^{-1}$ for GNR (**Figure 4b**). We thus conclude that GNRs are promising reinforce materials for PAN-based carbon nanofibers, and, given the recent dramatic improvements in the bottom-up GNR synthesis techniques,[21] suggest the utilization of various morphological degrees of freedom available in GNRs such as the GNR width and edge chirality as a promising future research direction. Here, we particularly note that the functionalization of GNR edges could be a potentially effective way to optimize PAN-GNR adhesion while preserving the good structural and electrical properties of basal planes.[22] This feature further differentiates GNRs from CNTs, in which chemical modifications inevitably involve the destruction of the sp$^2$ carbon network.[6, 17, 23] Overall, our simulation results provide justifications for recent experimental efforts to develop composite nanofibers containing CNTs with various diameters[24] or GNRs.[25-27]

## 3. Conclusion

In summary, carrying out combined quantum-mechanical DFT and classical FF MD simulations, we studied the PAN-CNT hybrid structure as a representative example of polymer-CNT composites. While one of the most promising applications of CNTs is to enhance polymer orientation and crystallization to achieve next-generation carbon fibers, atomistic understanding of the polymer-CNT interfaces still needs to be improved to successfully realize this goal. Based on DFT calculations, we first found that the relative orientation of the highly polar PAN nitrile group with respect to the CNT surface is the most critical factor that determines the PAN-CNT interface energetics and the lying-down PAN configurations are much more preferable than their standing-up counterparts. The CNT curvature was identified as another important factor, obtaining the largest PAN-CNT binding energy in the zero-curvature graphene limit. The quantum mechanical nature of our study revealed that the dominance of lying-down PAN configurations has significant implications for the PAN-CNT and PAN-PAN orderings: The charge transfer analysis revealed a linearly extended arrangement of the neighboring charge accumulation and depletion regions within CNT/graphene for the PAN backbone and nitrile side chain sides, respectively. Thus, we identified an electronic origin of the linear alignment of PAN on the CNT surface, a unique phenomenon observed more than a decade ago[8, 12] but whose mechanism so far remained unresolved.[13] Moreover, electrostatic potential profiles showed that the counterbalancing nature of charge accumulation and depletion regions and the resulting negligible net charge transfer at PAN-CNT interfaces allow them not to act as charged defects, enabling ordered long-range PAN-PAN stacking. Finally, performing large-scale FF MD simulations, we showed that the desirable linear PAN-CNT alignments arise even for large initial misorientation angles, and further demonstrated that graphene nanoribbons are a promising carbon nano-reinforcement candidate due to its zero-curvature nature and the possibility of edge functionalization. Given the enormous process parameters still need to be optimized, e.g. electrospinning[26] vs. gel spinning[27] in the preparation of PAN/GNR composite fibers and subsequent annealing conditions, a set of atomistic pictures established in this study will be valuable guidelines toward the development of next-generation carbon fibers.

## 4. Experimental Section

*DFT calculations*: In modeling the PAN polymer, we adopted the isotactic trimer units, which is the minimal model to describe the PAN backbone torsion (see Supporting Information Figure S2a). Having only end units, the dimer case is inadequate to describe the major interactions between the central PAN RUs and CNT. On the other hand, introducing more than two central PAN RUs induces the bending of the central units that result from the repulsion among the nitrile groups and makes the quantitative analysis complicated. We used a supercell with the dimension of 25 Å × 25 Å × Z Å, where Z is 17.22 Å for the armchair CNT (7 unit cells) and 17.04 Å for the zigzag CNT (4 unit cells). For the graphene case, we used a simulation box with the dimension of 16.97 Å × 16.79 Å × 20 Å (a rectangular cell based on 7 × 8 graphene unit cells). We carried out DFT calculations using the VASP package based on projected augmented wave scheme.[28] An energy cutoff of 400 eV was used, and only the Γ point of the Brillouin zone was sampled. In selecting the DFT exchange-correlation functional, we tested various approximations including LDA,[29] the generalized gradient approximation (GGA) within the Perdew-Burke-Ernzerhof parameterization,[30] and the GGA augmented by the D2,[31] D3,[15] and Tkatchenko-Scheffler[32] level van der Waals (vdW) corrections. For the test system of **L1** configuration at hollow site, as shown in Figure 1a, except the pure GGA that is well-known to give insufficient descriptions for dispersion interactions, we obtained consistent conclusions within different exchange-correlation energy functionals and carried out main DFT calculations within DFT-D3. Geometries of the PAN-CNT/graphene composite systems were optimized until the atomic forces decrease below 0.02 eV/Å while the CNT/graphene atomic structures were fixed to scan large conformational degrees of freedom. Full optimizations in several test cases resulted in only negligible changes in the results.

*FF MD simulations*: To carry out FF MD simulations, we used the LAMMPS (large-scale atomic/molecular massively parallel simulator) code.[33] To model the PAN polymer, we employed atactic 100 PAN RUs, which is long enough to probe the dynamics of the PAN-CNT interface. The atomic partial charges for PAN were calculated based on the Mulliken populations obtained from the DFT results produced with the Jaguar package[34] (the B3LYP exchange-correlation functional and a 6-31G** basis set. The PAN-CNT interaction was described based on the Lennard-Jones potential, which was carefully tested against DFT results (Supporting Information Table S5). We first prepared very long CNTs and GNR aligned along the z-axis. Then, the center of the stretched polymer chain was placed within 10Å displacement from the CNT surface. Various PAN-CNT configurations were explored by altering the orientation angle between the PAN chain and CNT. A large simulation box with the dimension of 200 Å × 200 Å × 300 Å was adopted to avoid artificial interactions of the PAN-CNT model with its periodic images. Newton's equations of motion were integrated using the velocity Verlet algorithm[35] at the time step of 1 fs. The van der Waals interactions were switched off between 12 Å to 13 Å. Electrostatics was treated with particle particle-mesh (PPPM) method,[36] using a short-range cutoff of 13 Å. To equilibrate the system, we first ran annealing simulations after energy minimizations. Then, canonical ensemble (NVT) simulation was performed for 2 (or 10 ns for multiple chains simulations). The temperature was maintained at 300K using the Nose-Hoover thermostat.[35] All MD simulations were repeated three times to check the validity (see Supporting Information **Table S7**).

## Acknowledgements

This work was supported by the Nano-Material Technology Development Program (Nos. 2016M3A7B4024133 and 2016M3A7B4909944), Basic Research Program (No. 2017R1A2B3009872), Global Frontier Program (No. 2013M3A6B1078881), and Basic Research Lab Program (No. 2016M3A7B4909944) of the National Research Foundation funded by the Ministry of Science and ICT of Korea.

## Corresponding Author Information

*Y.-H.K.: y.h.kim@kaist.ac.kr

## Keywords

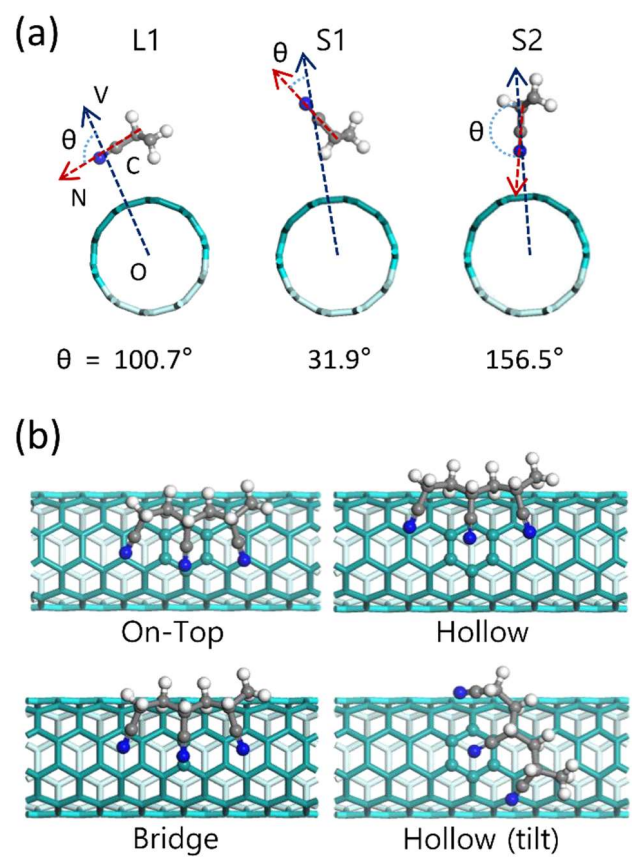

**Figure 1 |** Representative configurations of PAN on the surface of (5, 5) CNT defined based on a) the angle of PAN nitrile group in each RU along the CNT radial direction and b) the alignment of several RUs along the CNT axial direction. Quantifying the orientation PAN RUs in terms of the angle $\theta$ between the PAN nitrile group ($\overrightarrow{CN}$) with respect to the CNT radial direction ($\overrightarrow{OV}$), we considered one lying-down (**L1**) and two standing-up (**S1**, and **S2**) configurations. Additional considered configurations are described in Figure S1. The symbols of V and O indicate the mid-point of nitrile group and axis of CNT, respectively.

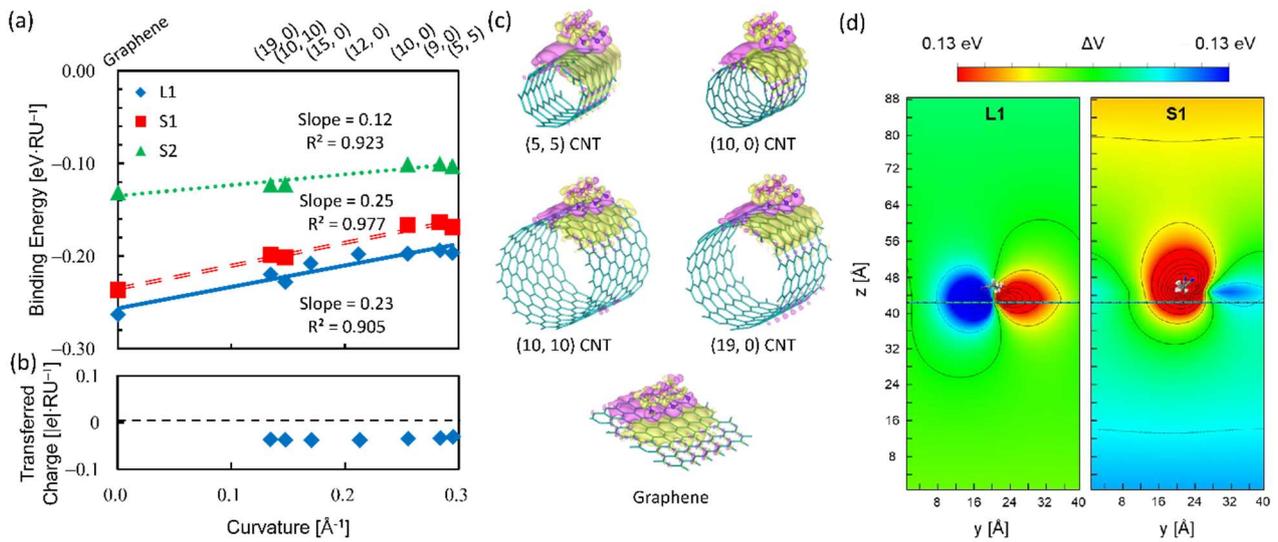

**Figure 2** | Energetic and electronic properties of PAN-CNT composite models calculated within DFT. a) Binding energies of all configurations as a function CNT diameter, including graphene as the zero-curvature limit. The dotted lines are shown as guide to the eyes. Corresponding b) charge transfers caculated within the Bader analysis scheme and c) charge density differences $\Delta\rho$ shown in real space. d) Electrostatic potentials induced by $\Delta\rho$ for the **L1** and **S1** PAN configurations on graphene shown along the graphene-normal direction (left and right panels, respectively). In c), purple and yellow surfaces depict the charge accumulation and depletion regions, respectively, and the isosurface level is $1.1 \times 10^{-4}$ e·Bohr$^{-3}$.

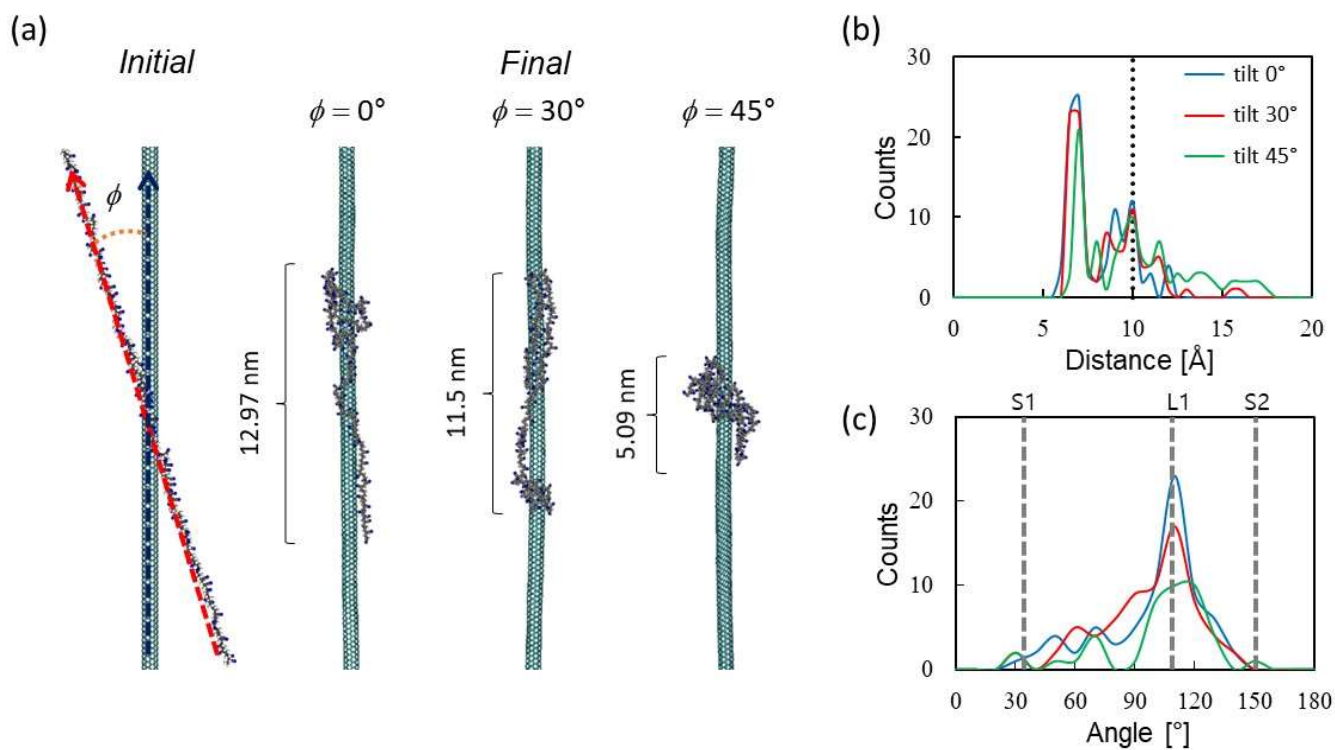

**Figure 3** | a) Initial and final conformations of a single PAN chain around a (5,5) CNT obtained from MD simulations started at three different initial axial-direction tilt angles $\phi$, which was defined as the angle between the end-to-end vector of stretched polymer (red dashed arrow) and the CNT axis (black dashed arrow). The PAN adsorption lengths measured along the CNT axis are given together. b) The cylindrical distributions of nitrile groups from the CNT axis. The dotted line indicates the approximate bondary, beyond which (10 Å, or 4.5 Å from the CNT surface) the PAN nitrile groups cannot be in contact with the surface of a (5,5) CNT. c) Distributions of readial-direction orientation angles $\theta$ of the nitrile groups located within the interface region obtained from the $\phi = 0°$ (blue), 30° (red), and 45° (green) MD runs. Gray dashed lines indicate the $\theta$ angles of the DFT-oopteimized **L1**, **S1**, and **S2** configurations shown in Figure 1a.

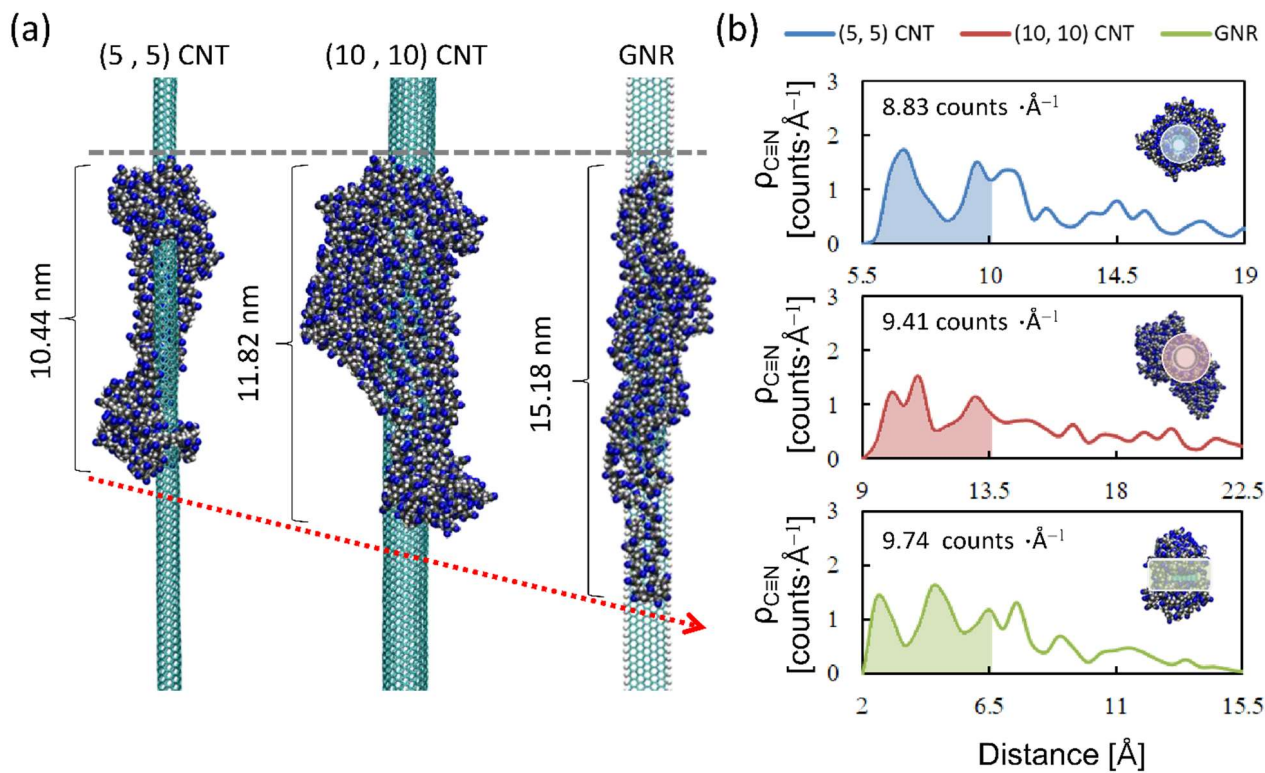

**Figure 4** | a) Conformations of multiple PAN chains around (5,5) CNT, (10,10) CNT, and GNR. The gray dashed line mark the starting point of PAN in three different cases. The red dotted arrow indicates the propensity of the increment of PAN adsorption length with the decrease in CNT curvature. b) The cylindrical distributions of PAN nitrile groups from the CNT axis in the (5, 5) CNT (upper panel), (10, 10) CNT (middle panel), and GNR (bottom panel) cases. The shaded area indicate the interface regions where PAN nitrile groups are in contact with the CNT or GNR surface (up to 4.5 Å along the surface normal direction). The insets show the top view of PAN-CNT composites with the interface regions indicated by shaded area.